\documentclass[aps,amssymb,amsmath,12pt,floatfix]{revtex4}
\usepackage{subfigure}
\usepackage{graphics}
\usepackage{amssymb,amsthm,amsmath,amsfonts,latexsym}
\usepackage{rotating}
\usepackage{algorithmic}
\usepackage[sort&compress]{natbib}

\newcommand{\algorithmicgpu}{\textbf{Beginning of GPU code section}}
\newcommand{\GPU}{\item[\algorithmicgpu]}
\newcommand{\algorithmicendgpu}{\textbf{End of GPU code section}}
\newcommand{\ENDGPU}{\item[\algorithmicendgpu]}

\begin{document}

\title{Efficient pseudo-random number generators for biomolecular simulations on graphics processors}

\author{A. Zhmurov$^{1,2}$, K. Rybnikov $^3$, Y. Kholodov$^1$ and V. Barsegov$^{2,1}$}
\thanks{Corresponding author; phone: 978-934-3661; fax: 978-934-3013;
Valeri\_Barsegov@uml.edu}
\affiliation{$^1$Moscow Institute of Physics and Technology, Dolgoprudnyi, Moscow region,\\
Russia, 141700, $^2$Department of Chemistry and $^3$Department of Mathematics, University 
of Massachusetts, Lowell, MA 01854}

\date{\today}


\begin{abstract}

\noindent Langevin Dynamics, Monte Carlo, and all-atom Molecular Dynamics simulations in implicit solvent, 
widely used to access the microscopic transitions in biomolecules, require a reliable source of random 
numbers. Here we present the two main approaches for implementation of random number generators 
(RNGs) on a GPU, which enable one to generate random numbers on the fly. In the one-RNG-per-thread 
approach, inherent in CPU-based calculations, one RNG produces a stream of random numbers in each 
thread of execution, whereas the one-RNG-for-all-threads approach builds on the ability of different 
threads to communicate, thus, sharing random seeds across the entire GPU device. We exemplify the use 
of these approaches through the development of Ran$2$, Hybrid Taus, and Lagged Fibonacci algorithms 
fully implemented on the GPU. As an application-based test of randomness, we carry out LD simulations of 
$N$ independent harmonic oscillators coupled to a stochastic thermostat. This model allows us to assess 
statistical quality of random numbers by comparing the simulation output with the exact 
results that would be obtained with truly random numbers. We also profile the performance of these generators 
in terms of the computational time, memory usage, and the speedup factor (CPU/GPU time).

\end{abstract}
\maketitle


\newpage

\section{Introduction}
\label{Introduction}

Over the last few years, graphics processors have evolved into highly parallel, multithreaded computing 
devices. Graphics Processing Units (GPUs) are now emerging as an alternative programming platform 
that provides high raw computational power for scientific applications \cite{Schulten07,Pande09,Anderson08,
Meel08,Harvey09,Davis09,Anderson07,Yang07}. Because GPUs implement Single Instruction Multiple Data (SIMD) 
architecture with reduced cache and flow control for a group of computational cores, most of a GPU 
device form computational units dedicated to actual calculations. The computational efficiency of contemporary 
GPUs can reach striking $1$ TFlops for a single chip \cite{CUDAPG}, the number not yet accessible even for 
most up-to-date CPUs. With introduction of CUDA (Compute Unified Device Architecture) by NVIDIA (a dialect 
of C and C++ programming languages) \cite{CUDAPG,CUDABP}, GPUs have become capable of performing compute-intensive 
scientific calculations. Because the GPU-based calculations are $10$$-$$50$ times faster than some of the 
heavily tuned CPU-based methods, GPUs are being used as performance accelerators in a variety of applications 
\cite{Schulten07,Pande09,Anderson07,Yang07}.

The GPU-based calculations can be performed concurrently on many computational cores, called Arithmetic Logic 
Units (ALUs), that are grouped into multiprocessors, each with its own flow control and cache units. For 
example, in contemporary graphics cards from NVIDIA each multiprocessor contains up to eight $\sim$$1.3GHz$ 
ALUs and $14$$-$$16KB$ of cache ($8KB$ of constant memory cache and $6$$-$$8KB$ of the global memory cache 
when accessed through texture references). The number of multiprocessors per GPU can reach $30$ on the most 
up-to-date graphics cards (Tesla C1060 or GeForce GTX 285), thus, bringing the total number of ALUs to $240$ 
per chip. Due to the inherently parallel nature of the GPU-based calculations, achieving optimal performance 
on the GPU mandates that a computational task be divided into many independent threads of execution 
that run in parallel performing the same operation but on different data sets. Although each graphics card 
has its own global memory with $\sim$$10$ times larger bandwidth compared to DRAM on a CPU, the number of 
memory invocations (per ALU) should be minimized to optimize the GPU performance. Hence, the computational
task should be compute-intensive so that, most of the time, the GPU is busy performing computations rather 
than reading and writing data \cite{CUDAPG,CUDABP}. This makes a classical $N$ body problem that would be 
difficult or impossible to solve exactly into a prime candidate for the numerical implementation on the GPU. 

Computer simulations of a system of $N$ particles, e.g., Langevin Dynamics (LD), Monte Carlo (MC), and 
Molecular Dynamics (MD) simulations, are among many applications that can be implemented on the GPU. In LD 
and MD simulations, atomic interactions are described by the same potential energy function (force field) 
applied to all particles in the system. Hence, there is a direct mapping between the SIMD architecture of the 
GPU (hardware) and numerical routines (software) used to follow a trajectory of the system under the study in 
real time. In a sense, a ``single instruction'', i.e. calculation of the potential energy terms or evaluation 
of forces and random forces, or numerical integration of the equation(s) of motion, is executed on ``multiple 
data'' sets (for all particles) in order to describe the dynamics of the whole system. For example, in MD 
simulations of biomolecules in implicit solvent (water) \cite{CHARMM83,Caflisch08}, the dynamics of the $i$-th 
particle are governed by the equations of motion for the particle position, 
$d{\bf{R}}_i/dt$$=$${\bf{V}}_i$ (${\bf{R}}_i$$=$$\{R_{i,x},R_{i,y},R_{i,z}\}$), and velocity, 
$m_i$$d{\bf{V}}_i/dt$$=$$\xi$${\bf{V}}_i$$+$${\bf{f}}({\bf{R}}_i)$$+$${\bf{G}}_i(t)$ 
(${\bf{V}}_i$$=$$\{V_{i,x},V_{i,y},V_{i,z}\}$), where $m_i$ is the particle mass, $\xi$ is the friction 
coefficient, ${\bf{f}}({\bf{R}}_i)=-\partial{\bf{U}}/\partial{\bf{R}}_i$ is the molecular force exerted 
on the $i$-th particle due to the potential energy ${\bf{U}}$$=$${\bf{U}}({\bf{R_1,R_2,\ldots,R_N}})$, and 
${\bf{G}}_i(t)$$=$$\{G_{i,x},G_{i,y},G_{i,z}\}$ is the Gaussian random force with the first moment 
$\langle {\bf{G}}_i(t)\rangle$$=$$0$ and the two-point correlation function 
$\langle {\bf{G}}_i(t){\bf{G}}_j(t')\rangle=2k_BT$$\xi\delta_{ij}\delta$$(t-t')$ ($i,j$$=$$1,2,\ldots,N$) 
\cite{Risken}. In LD simulations of proteins, the dynamics of the $i$-th $C_{\alpha}$-particle is governed 
by the Langevin equation for ${\bf{R}}_i$, $\xi$$d{\bf{R}}_i/dt$$=$${\bf{f}}({\bf{R}}_i)$$+$${\bf{G}}_i(t)$ 
\cite{Doi}. These equations of motion are solved numerically over many iterations of the simulation algorithm. 

Since in MD simulations in implicit water and in LD simulations, the effect of solvent molecules is described 
implicitly, these methods require a reliable source of $3$$N$ normally distributed random numbers, 
$g_{i,\alpha}$ ($i$$=$$1,2,\ldots,N$) to compute the components of the Gaussian random force 
$G_{i,\alpha}$$=$$g_{i,\alpha}$$\sqrt{2k_BT\xi\Delta t}$, where $\Delta$$t$ is the time step ($\alpha$$=$$x$, 
$y$, and $z$). In MC simulations, the results of multiple independent trials, each driven by some random process, 
are combined to extract the average answer. A pseudo-random number generator, or algorithmic RNG, must have a 
long period and must meet the conflicting goals of being fast while also providing a large amount of random numbers 
of proven statistical quality \cite{Lecuyer07}. An RNG produces a deterministic sequence of random numbers, 
$u_{i}$, that are supposed to imitate realizations of independent uniform random variables from the interval 
$(0,1)$, i.e., i.i.d. $U(0,1)$. This sequence ($u_{i}$) is translated into the sequence of normally distributed 
random variables ($g_{i}$) using the ziggurat method \cite{Marsaglia00} or the polar method \cite{Marsaglia64}, 
or the Box-Mueller transformation \cite{Box58}. There is an extensive body of literature devoted 
to random number generation on the CPU \cite{Recipes}. Yet, due to fundamental differences 
in processor and memory architecture of CPU and GPU devices, the CPU-based methods cannot be easily translated 
from the CPU to the GPU. While there exist stand-alone implementations of good quality RNGs on the GPU, to 
fully utilize computational resources of the GPU in molecular simulations an RNG should be incorporated into 
the main simulation program. This enables a developer to minimize read/write calls associated with invocation 
of the relatively slow GPU global memory, and to generate streams of random numbers using fast GPU shared memory. 

We present a novel methodology for generating pseudo-random numbers on a GPU on the fly, i.e. at each step 
of a simulation run. This methodology can be used in the development of the GPU-based implementations of MD 
simulations in implicit solvent, and in LD and MC simulations. We focus on the Linear Congruential Generator 
(LCG), and the Ran$2$, Hybrid Taus, and Lagged Fibonacci algorithms reviewed in the next Section. These 
algorithms are used in Section \ref{Implementation} to describe the one-RNG-per-thread approach and the 
one-RNG-for-all-threads approach for random number generation on the GPU. In the one-RNG-per-thread setting, 
one RNG is assigned for each computational thread (for each particle), a procedure commonly used in the 
CPU-based calculations. The one-RNG-for-all-threads method utilizes the ability of different threads to 
communicate across the entire GPU device (pseudocodes are given in the Appendices). We test the performance 
of GPU-based implementations of these generators in Section \ref{ApplicationTest}, where we present application-based 
assessment of their statistical qualities using Langevin simulations of $N$ independent Brownian particles evolving 
on the harmonic potential. We profile these generators in terms of the computational time and memory usage 
for varying system size $N$. The main results are discussed in Section \ref{Discussion}, where we provide
recommendations on the use of RNG algorithms.


\section{Pseudorandom number generators}
\label{Generators}


\subsection{Overview}
\label{Generators.Overview} 

There are three types of random numbers generators: true or hardware random numbers generators, and
software-based quasi-random numbers generators and pseudo-random numbers generators (RNGs) \cite{GPUGems3}. 
In this paper we focus on algorithmic RNGs - the most common type of deterministic random number generators. 
Because an RNG produces a sequence of random numbers in a purely deterministic fashion, a good quality 
RNG should have a long period and should pass some stringent statistical tests for uniformity and independence. 
MD simulations of biomolecules in implicit water and LD simulations of proteins use normally distributed random 
forces to emulate stochastic kicks from the solvent molecules. To generate the distribution of random forces, a 
common approach is to convert the uniformly distributed random variates into the Gaussian distributed 
random variates using a particular transformation. In this paper, we adopt the most commonly used 
Box-Mueller transformation \cite{Box58}.

There are three main requirements for a numerical implementation of an RNG: ($1$) good statistical properties, 
($2$) high computational speed, and ($3$) low memory usage. A deterministic sequence of random numbers comes 
eventually to a starting point, i.e. to the initial set of random seeds $u_{n+p}$$=$$u_n$ \cite{Barreira06}. 
This mandates that an RNG should have a long period $p$. For example, a simulation run might use $10^{12}$ random 
numbers, in which case the period must far exceed $10^{12}$. Once an RNG has been selected and implemented, it 
must also be tested empirically for randomness, i.e., for the uniformity of distribution and for the independence 
\cite{Lecuyer07}. In addition, it must also pass application-based tests of randomness that offer exact solutions 
to the test applications. Using random numbers of poor statistical quality in molecular simulations might result 
in insufficient sampling, unphysical correlations or even patterns \cite{Selke93,Grassberger93}, and unrealistic 
results, which leads to errors in practical applications \cite{Ferrenberg92}. Some of the statistical tests 
of randomness are accumulated in the DIEHARD test suite and in the TestU01 library \cite{Lecuyer07,Marsaglia96,
Mascagni00,Soto99}. In numerical implementations, a good quality RNG should also be computationally efficient 
so that random number generation does not become a major bottleneck. For example, in LD simulations of proteins 
on a GPU, one can obtain long $0.1s$ trajectories over as many as $10^{10}$ iterations. Hence, to simulate 
one trajectory for a system of $10^3$ particles requires total of $\sim$$10^{13}$ random numbers. The requirement 
of low memory usage is also important as modern graphics processors have low on-chip memory, $\sim20KB$ per 
multiprocessor, compared to $\sim$$2MB$ memory on the CPU. Hence, an efficient RNG algorithm must use limited 
working area without invoking the slow GPU global memory. 

Typically, a fast RNG employs simple logic and a few state variables to store its current state, but this may 
harm statistical properties of the random numbers produced. On the other hand, using more sophisticated
algorithms with many arithmetic operations or combining several generators into a hybrid generator allows
to improve statistics, but these generators are slower and use more memory. Hence, a choice of RNG is
determined primarily by specific needs of a particular application, including statistical characteristics of
random numbers, and GPU capabilities. In this paper, we focus on the most widely used algorithms: Linear Congruential 
Generator (LCG) \cite{Recipes}, and the Ran$2$ \cite{Recipes}, Hybrid Taus generator \cite{GPUGems3,Tausworthe65,
Lecuyer96,Recipes} and Lagged Fibonacci algorithm \cite{Recipes,Mascagni04,Brent92}. LCG can be used in performance 
benchmarks since it employs a very fast algorithm. Ran$2$ is a standard choice for MD simulations of biomolecules 
in implicit water and in LD simulations of proteins due to its long period ($p$$>$$2$$\times$$10^{18}$), good 
statistical quality, and high computational performance on the CPU. However, Ran$2$ requires large amount of 
on-chip GPU local memory and global memory to store its current state. Hybrid Taus is an example of how several 
simple algorithms can be combined to improve statistical characteristics of random numbers. It scores better in 
terms of the computational speed on the GPU than KISS, the best known combined generator \cite{Marsaglia99}, and 
its long period ($p$$>$$2$$\times$$10^{36}$) makes it a good choice for molecular simulations on the GPU. 
Lagged Fibonacci employs very simple logic while producing random numbers of high statistical 
quality \cite{Lecuyer07,Brent92}. It is commonly used in distributed MC simulations, and it can also be 
utilized in GPU-based computations. Here, we briefly review the LCG, Ran$2$, Hybrid Taus, and 
Lagged Fibonacci algorithms.


\subsection{Linear Congruential Generator} 
\label{Generators.LCG}

The Linear Congruential Generators (LCGs) are the classic and most popular class of generators, which use 
a transitional formula, 
\begin{equation}\label{eq1}
x_{n}=(ax_{n-1} + c)\:\mathrm{mod}\:m, 
\end{equation}
where $m$ is the maximum period, and $a$$=$$1664525$ and $c$$=$$1013904223$ are constant parameters \cite{Recipes}. 
To produce a uniformly distributed random number, $x_{n}$ is divided by $2^{32}$. Assuming a 32-bit 
integer, the maximum period can be at most $p$$=$$2^{32}$, which is far too low.  LCGs also have known statistical 
flows \cite{Lecuyer07}. If $m$$=$$2^{32}$, one can neglect mod $m$ operation as the returned value is low-order 
$32$ bits of the true $64$-bit product. Then, the transitional formula reads $x_{n}$$=$$ax_{n-1}$$+$$c$, 
which is the so-called Quick and Dirty (or ranqd$2$) generator (simplified LCG generator). Quick and Dirty LCG 
is a very fast generator as it takes only a single multiplication and a single addition to produce a random 
number, and it uses a single integer to describe its current state. Due to low memory usage, Quick and Dirty 
LCG can be used to benchmark GPU-based implementations of software packages at the development stage.


\subsection{Ran$2$}
\label{Generators.Ran2} 

Ran$2$, one of the most popular RNGs, combines two LCGs and employs randomization using some shuffling procedure 
\cite{Recipes}. Ran$2$ has a long period and provides random numbers of very good statistical 
properties \cite{Lecuyer07}. In fact, Ran$2$ is one of a very few generators that does not fail a single known 
statistical test. It is reasonably fast, but there are several features that make Ran$2$ less attractive for GPU-based 
computations. First, the algorithm involves long integer arithmetic ($64$-bit logic) - a computational 
bottleneck for contemporary GPUs. Secondly, it requires a large amount of memory to store its current state that
needs to be updated at each step. This involves a large number of memory calls, which may, potentially, slow 
down the computational speed on the low cache GPU.


\subsection{Hybrid Taus}
\label{Generators.HybridTaus}

Hybrid Taus \cite{GPUGems3} is a combined generator that uses LCG and Tausworthe algorithms. Tausworthe taus$88$ 
is a fast equidistributed modulo $2$ generator \cite{Tausworthe65,Lecuyer96}, which produces random numbers by 
generating a sequence of bits from a linear recurrence modulo $2$, and forming the resulting number by taking a 
block of successive bits. In the space of binary vectors, the $n$-th element of a vector is constructed using 
the linear transformation,
\begin{equation}\label{eq2} 
y_n=a_1y_{n-1}+a_2y_{n-2}+\ldots a_k y_{n-k},
\end{equation} 
where $a_n$ are constant coefficients. Given initial values, $y_0,y_1,\ldots y_{n-1}$, the $n$-th random integer 
is obtained as $x_n$$=$$\sum_{j=1}^L$$y_{ns+j-1}$$2^{-j}$, where $s$ is a positive integers and $L$$=$$32$ is the 
integer size (machine word size). Computing $x_n$ involves performing $s$ steps of the recurrence, which might 
be costly computationally. Fast implementation can be achieved for a certain choice of parameters. When 
$a_k$$=$$a_q$$=$$a_0$$=$$1$, where $0$$<$$2q$$<k$ and $a_n$$=$$0$ for $0$$<$$s$$\leq$$k$$-$$q$$<$$k$$\leq$$L$, the 
algorithm can be simplified to a series of binary operations \cite{Lecuyer96}. Statistical characteristics of 
random numbers produced using taus$88$ alone are poor, but combining taus$88$ with LCG removes all the statistical 
defects \cite{GPUGems3}. In general, statistical properties of a combined generator are better than those of 
components. When periods of all components are co-prime numbers, a period of a combined generator is the product 
of periods of all components. A similar approach is used in the KISS generator, which combines LCG, Tausworhe 
generator, and a pair of multiple-with-carry generators \cite{Marsaglia99}. However, multiple 32-bit multiplications, 
used in KISS, may harm performance on the GPU. The period is the lowest common multiplier of the periods of three 
Tausworthe steps and one LCG. We used parameters that result in periods of $p_1$$=$$2^{31}$$-$$1$, $p_2$$=$$2^{30}$$-$$1$, 
and $p_3$$=$$2^{28}$$-$$1$ for the Tausworthe generators and a period of $p_4$$=$$2^{32}$ for the LCG, which makes 
the period of the combined generator equal $\sim$$2^{121}$$>$$10^{36}$. Hybrid Taus uses small memory area since 
only four integers are needed to store its current state.


\subsection{Lagged Fibonacci}
\label{Generators.LaggedFibonacci}

The Lagged Fibonacci algorithm is defined by the recursive relation,
\begin{equation}\label{eq3}
x_n=f(x_{n-sl},x_{n-ll}) \;\mathrm{mod}\; m,
\end{equation}
where $sl$ and $ll$ are the short and long lags, respectively ($ll$$>$$sl$), $m$ defines the maximum period and $f$ 
is a function that takes two integers $x_{n-sl}$ and $x_{n-ll}$ to produce integer $x_n$. The most commonly used 
functions are multiplication, $f(x_{n-sl},x_{n-ll})$$=$$x_{n-sl}$$*$$x_{n-ll}$ (multiplicative Lagged Fibonacci), 
and addition, $f(x_{n-sl},x_{n-ll})$$=$$x_{n-sl}$$+$$x_{n-ll}$ (additive Lagged Fibonacci). Random numbers are 
generated from the initial set of $ll$ integer seeds, and to achieve the maximum period $\sim 2^{ll-1}$$\times$$m$
the long lag $ll$ should be set equal the base of a Mersenne exponent, and the short lag $sl$ should be taken so 
that the characteristic polynomial $x^{ll}$$+$$x^{sl}$$+$$1$ is primitive. In addition, $sl$ should not be too 
small or too close to $ll$; it is recommended that $sl$$\approx$$\rho$$\times$$ll$, where $\rho$$\approx$$0.618$ 
\cite{Brent92}. When single precision arithmetic is used, the $\mathrm{mod}\; m$ operation can be omitted by setting 
$m$$=$$2^{32}$ (more on selection of parameters can be found in Ref.~\cite{Brent92,Brent03}). We employed the additive 
Lagged Fibonacci RNG, which generates floating point variates directly, without the usual floating of random integers. 
Also, $sl$ and $ll$ can be taken to be very large, which improves statistical quality of the generator.


\section{GPU-based implementation of LCG, Ran$2$, Hybrid Taus, and Lagged Fibonacci algorithms}
\label{Implementation}


\subsection{Basic ideas}
\label{Implementations.BasicIdeas}

The main feature that makes GPUs computationally efficient is their many-thread architecture, i.e. calculations 
are performed on a GPU using threads working in parallel. Hence, in molecular simulations of an $N$ body system on 
a GPU an RNG should produce independent random numbers simultaneously for all particles. One possibility is to 
have random numbers pre-generated on a CPU or on a GPU, and then use these numbers in simulations. However, this 
requires a large amount of memory allocated for an RNG. For example, for a system of $10^4$ particles in three 
dimensions, $3$$\times$$10^4$ random numbers are needed at each simulation step. If these numbers are pre-generated, 
say, for every $100$$-$$1000$ steps, it requires $3$$\times$$10^{6}$$-$$3$$\times$$10^{7}$ random numbers to be 
stored on the GPU, which takes $12$$-$$120MB$ of memory. This might be significant for graphics cards with limited 
memory, e.g., GeForce GTX 200 series (from NVIDIA) with $\sim$$1GB$ of memory. Another approach is to build an RNG 
into the main simulation kernel. This allows one to achieve top performance for an RNG by maximizing the 
amount of computations on a GPU while also minimizing the number of calls of the GPU global memory (read/write 
operations). In addition, to fully utilize the GPU resources, the total number of threads should be $\sim$$10$-times 
larger than the number of computational cores, so that none of the cores awaits for the others to complete their
tasks. 

To develop parallelized implementations of several different RNGs on the GPU, we employ cycle division paradigm 
\cite{Mascagni04}. The idea is to partition a single RNG sequence, which can be viewed as a periodic circle of 
random numbers, among many computational threads running concurrently across the entire GPU device, each producing 
a stream of random numbers. Since most RNG algorithms are based on sequential transformations of the current state, 
including LCG, Hybrid Taus and Ran$2$, the most common way of partitioning the sequence is to provide each thread 
with different seeds while also separating the threads along the sequence to avoid possible inter-stream
correlations. This is the basis of the one-RNG-per-thread approach (Fig.~1). Also, there exist RNG, 
e.g., Mersenne Twister and Lagged Fibonacci algorithms, that allow one to leap ahead in the sequence to produce 
the $(n+1)$ random number without first computing the $n$-th number \cite{Marsaglia99,Lecuyer93,Mascagni04}. The 
leap size, which, in general, depends on parameters of an RNG, can be adjusted to the number of threads (equal 
the number of particles $N$), or multiples of $N$ ($M$$\times$$N$). Then, all $N$ random numbers can be obtained
simultaneously, i.e. the $j$-th thread produces numbers $j$, $j$$+$$N$, $j$$+$$2N$$\ldots$, etc. Note that at 
the end of each simulation step, threads must be syncronized so that the current RNG state is properly updated. 
The same RNG state is used by all threads, each updating just one elements of the state. We refer to this 
as the one-RNG-for-all-threads approach (Fig.~1). In what follows, we describe these approaches in more detail.


\subsection{One-RNG-per-thread approach}
\label{Implementation.ManyRNGs} 

The idea is to run the same RNG algorithm in many threads, where all RNGs generate different subsequences 
of the same sequence of random numbers using the same algorithm, but starting from different initial seeds. 
First, a CPU generates $N$ sets of random seeds (one for each RNG) and passes them to the GPU global memory 
(Fig.~2). To exclude correlations, these sets should come from an independent sequence of random numbers, or  
should be generated using different RNG algorithms on the CPU. In a simulation run, each thread on the 
GPU reads its random seeds from the GPU global memory and copies them to the GPU local (per thread) memory 
or shared (per thread block) memory. Then, each RNG can generate as many random numbers as needed, without using 
the slow GPU global memory. At the end of a simulation step, each RNG saves its current state to the global 
memory and frees shared memory. Since each thread has its own RNG, there is no need for threads synchronization; 
however, when particles interact threads must be synchronized. In molecular simulations of a system of $N$ 
particles, $4$$N$ uniformly distributed random variates are needed at each step, and arrays of initial 
seeds and the current state should be arranged for coalescent memory read to speedup the global memory access. 
In the one-RNG-per-thread setting, an RNG should be very light in terms of memory usage. Small size of 
on-chip memory can be insufficient to store the current state of an RNG that is based on a complex algorithm. 
These restrictions make it difficult to use simple RNG algorithms, especially when statistical properties of 
random numbers become an issue.

In the one-RNG-per-thread approach, the amount of memory required to store the current state of a generator 
is proportional to the number of threads (number of particles $N$). Hence a significant amount of memory has 
to be allocated for all RNGs to describe the dynamics for a large system. For example, LCG uses one integer 
seed to store its current state, which takes $4$ bytes per thread (per generator) or $\sim$$4MB$ of memory 
for $10^6$ threads, whereas Hybrid Taus uses $4$ integers, i.e. $16MB$ of memory. These are acceptable numbers, 
given hundreds of megabytes of the GPU memory. By contrast, Ran$2$ uses $35$ long integers and a total of $280$ 
bytes per thread, or $\sim$$280MB$ of memory (for $10^6$ threads). As a result, not all seeds can be stored 
in on-chip (local or shared) memory ($\sim$$16KB$), and the Ran$2$ RNG has to access the GPU global memory to 
read and update its current state. In addition, less memory becomes accessible to other computational routines. 
This might prevent using Ran$2$ in the simulations of large systems on some graphics cards, including GeForce 
GTX 280 and GTX 295 (NVIDIA), with $768MB$ of global memory (per GPU). However, this is not an issue when using 
high end graphics cards, such as Tesla C1060 with $4GB$ of global memory. In this paper, we utilized the 
one-RNG-per-thread approach to develop the GPU-based implementations of the Hybrid Taus and Ran$2$ algorithms 
(pseudocodes are presented in Appendix A).


\subsection{One-RNG-for-all-threads approach}
\label{Implementation.OneRNG}

Within the one-RNG-for-all-threads approach, one can use a single RNG by allowing all computational 
threads to share the state of a generator. This approach can be adapted to RNG algorithms that are based on 
the recursive transformations, i.e., $x_n$$=$$f(y_{n-r},y_{n-r+1},\ldots y_{n-k})$, where $r$ is the recurrence 
degree and $k$$>$$r$ is a constant parameter. This transformation allows one to obtain a random number at 
the $n$-th step from the state variables generated at the previous steps $n$$-$$r$, $n$$-$$r$$+$$1$, $\ldots$, 
$n$$-$$k$. If a sequence of random numbers is obtained simultaneously in $N$ threads, each generating just one 
random number at each step, then total of $N$ random numbers are produced. Then, given $k$$>$$N$, all the elements 
of the transformation have been obtained in the previous steps, in which case they can be accessed without 
threads synchronization. One of the algorithms that can be implemented on the GPU using the one-RNG-for-all-threads 
approach is Lagged Fibonacci (Fig.~3) \cite{Lecuyer93}. When one random number is computed in each thread and 
when $sl$$>$$N$ and $ll$$-$$sl$$>$$N$ (Section \ref{Generators.LaggedFibonacci}), $N$ random numbers can be 
obtained simultaneously on the GPU without threads synchronization. 

To initialize the Lagged Fibonacci RNG on the GPU, $ll$ integers are allocated on the CPU using initial 
seeds. Each thread then reads two integers from this sequence, which correspond to the long lag $ll$ and 
the short lag $sl$, generates the resulting integer, and saves it to the location in the GPU global memory, 
which corresponds to the long lag. Setting $sl$$>$$N$ and $ll$$-$$sl$$>$$N$ guarantees that the same position 
in the array of integers (current state variables) will not be accessed by different threads at the same time. 
The moving window of $N$ random numbers, updated by $N$ threads at each step, is circling along the array of 
state variables, leaping forward by $N$ positions (at each step). Importantly, a period of the Lagged Fibonacci 
generator, $p$$\sim$$2^{ll+31}$, can be adjusted to the system size by assigning large values to $sl$ and 
$ll$, so that $p$$\gg$$N$$\times$$S$, where $S$ is the total number of simulation steps. Changing $ll$ 
and $sl$ does not influence the execution time, but affects the size of the array of state variables, which 
scales linearly with $ll$ - the amount of integers stored in the GPU global memory. Large $ll$ is not an issue 
even when $ll$$\sim$$10^6$, which corresponds to $\sim$$4MB$ of the GPU global memory (pseudocode for the 
Lagged Fibonacci RNG is presented in Appendix B). Note, that the GPU-based implementation of the Lagged Fibonacci 
algorithm using the one-RNG-per-thread approach requires to store $N$ independent RNG states of size $ll$, i.e. 
$N$ times larger memory.


\section{Application-based test of randomness: Ornstein-Uhlenbeck process}
\label{ApplicationTest}

To assess the computational and statistical performance of the LCG, Ran$2$, Hybrid Taus, and Lagged 
Fibonacci algorithms in molecular simulations, we carried out Langevin simulations of $N$ independent one-dimensional 
harmonic oscillators in a stochastic thermostat, fully implemented on the GPU. Each particle evolves 
on the harmonic potential, $V(R_i)$$=$$k_{sp}R_i^2/2$, where $R_i$ is the $i$-th particle position and $k_{sp}$ 
is the spring constant. We employed this analytically tractable model from statistical physics to compare the 
results of simulations with the theoretical results that would be obtained with truly random numbers. In the 
test simulations, we used NVIDIA graphics card GeForce GTX 295, which has two processing units (GPUs), each with 
$30$ multiprocessors (total of $240$ ALUs) \cite{CUDAPG} and $768MB$ of global memory. 

The Langevin equations of motion in the overdamped limit, 
\begin{equation}\label{eq4}
\xi {{dR_i} \over {dt}}=-{{\partial{V(R_1,R_2,\ldots,R_N)} } \over {\partial{R_i} }}+G_i(t), 
\end{equation}
were integrated numerically using the first-order integration scheme (in powers of the integration time step 
$\Delta$$t$) \cite{Ermak78}, 
\begin{equation}\label{eq5}
R_i(t+\Delta t)=R_i(t)+f(R_i(t))\Delta t/\xi+g_i(t)\sqrt{2k_BT\xi\Delta t},
\end{equation}
where $f(R_i)$$=$$-$$(\partial{V(R_1,R_2,\ldots,R_N)}/\partial{R_i})$ is the force acting on the $i$-th oscillator
\cite{Hummer03,Valeri06,Ruxandra07}. In Eq. (\ref{eq5}), $g_i$ are the Gaussian distributed random variates (with 
zero mean and unit variance), which are transformed into the random forces $G_i(t)$$=$$g_i(t)$$\sqrt{2k_BT\xi \Delta t}$. 
Langevin dynamics in the overdamped limit (Eqs. (\ref{eq4}) and (\ref{eq5})) are widely used in the simulations of 
biomolecules \cite{Tozzini05,Clementi08,Ruxandra06,Ruxandra07,Ruxandra08,Dmitri97,Valeri06}. Numerical values of 
the constant parameters for the LCG, Ran$2$, Hybrid Taus, and Lagged Fibonacci algorithms can be found, respectively, 
in Section II \cite{Lecuyer07}, in Ref.~\cite{Recipes}, in Appendix A, and in Table I. 

We employed the one-RNG-per-thread approach to develop the GPU-based implementations of the LCG, Ran$2$, and 
Hybrid Taus algorithms, and used the one-RNG-for-all-threads approach for the Lagged Fibonacci RNG. These 
implementations have been incorporated into the LD simulation program written in CUDA. Numerical algorithms for 
the GPU-based implementation of LD simulations of biomolecules, which involves evaluation of the potential energy, 
calculation of forces, and numerical integration of the Langevin equations of motion, will be presented in a separate 
publication (A. Zhmurov, R. I. Dima, Y. Kholodov, and V. Barsegov, submitted to {\em J. Chem. Theory and Comput.}) In our 
implementation, each computational thread generates one trajectory for each particle, and we used $64$ threads in a 
thread block. Numerical calculations for $N$$=$$10^4$ particles were carried out with the time step $\Delta$$t$$=$$1ps$, 
starting from the initial position $R_0$$=$$10nm$, and using $k_{sp}$$=$$0.01pN/nm$, $T=300K$, and $D=0.25nm^2/ns$. 
Soft harmonic spring ($0.01pN/nm$) allowed us to generate long $1 ms$ trajectories over $10^9$ steps. We analyzed 
the average position $\langle R(t)\rangle$ and two-point correlation function $\langle R(t)R(0)\rangle$, obtained 
from simulations, and have compared these quantities with their exact counterparts \cite{Doi,Risken}, 
$\langle R(t)\rangle $$=$$R_i(0)$$\exp{[-t/\tau]}$ and $\langle R(t)R(0)\rangle $$=$$(k_BT/k_{sp})$$\exp{[-t/\tau]}$, 
respectively, where $\tau$$=$$\xi/k_{sp}$ is the characteristic time. All RNGs describe well the exact Brownian dynamics 
except for LCG (Fig.~4). Both $\langle R(t)\rangle$ and $\langle R(t)R(0)\rangle$, obtained using Ran$2$, Hybrid Taus, and 
Lagged Fibonacci, practically collapse on the theoretical curve of these quantities. By contrast, using LCG results 
in repeated patters of $\langle R(t)\rangle$ and unphysical correlations in $\langle R(t)R(0)\rangle$ (Fig.~4). At 
longer times, $\langle R(t)\rangle$ and $\langle R(t)R(0)\rangle$, obtained from simulations, deviate somewhat from 
the theoretical curves due to a soft harmonic spring and insufficient sampling (Fig.~4). 

In biomolecular simulations on a GPU, a large memory area should be allocated to store parameters of the 
force field, Verlet lists, interparticle distances, etc., and the memory demand scales with the system size as 
$\sim$$N^2$. In contemporary graphics cards, the amount of global memory is low, and each memory access takes 
$\sim$$300$ clock cycles. The number of memory calls scales linearly with the amount of random numbers produced. 
Because the computational speed even of a fast RNG is determined mostly by the number of global memory calls, 
multiple reads and writes from and to the GPU global memory can prolong significantly the computational time. We 
profiled the LCG, and the Ran$2$, Hybrid Taus, and Lagged Fibonacci RNGs in terms of the number of global memory 
calls per simulation step. These generators use, respectively, $1$, $40$, $4$, and $\sim$$3$ random seeds per 
thread (the state size for Lagged Fibonacci depends on the choice of parameters $ll$ and $sl$). In our implementation, 
the LCG, and the Hybrid Taus and Lagged Fibonacci RNGs use $4$$-$$16$ bytes of memory per thread, which is quite 
reasonable even for large system size $N$$=$$10^6$. However, Ran$2$ requires $280$ bytes per thread which is significant 
for a large system (Table II). Ran$2$ has large state size, and saving and updating its current state using the GPU local 
or shared memory is not efficient computationally. Ran$2$ uses long $64$-bit variables, which doubles the amount of 
data, and requires $4$ read and $4$ write calls ($7$ read and $7$ write memory calls are needed to generate $4$ 
random numbers). The Hybrid Taus RNG uses the GPU global memory only when it is initialized, and when it updates its 
current state. Since it uses $4$ state variables, $4$ read and $4$ write memory calls per thread are required 
irrespectively of the amount of random numbers (Table II). The Lagged Fibonacci RNG uses $2$ random seeds, which results 
in $2$ read and $1$ write memory calls per random number, and $8$ read and $4$ write calls for four random numbers
(Table~II).

To benchmark the computational efficiency of the LCG, and the Ran$2$, Hybrid Taus, and Lagged Fibonacci RNGs, we 
carried out LD simulations of $N$ three-dimensional harmonic oscillators in a stochastic thermostat. For each $N$, 
we generated one simulation run over $n$$=$$10^3$ steps. All $N$ threads have been synchronized at the end of each 
step to emulate an LD simulation run of a biomolecule on a GPU. The execution time and memory usage are displayed 
in Fig.~5. Ran$2$ is the most demanding generator: the use of Ran$2$ in LD simulations of a system of $10^4$ particles 
adds extra $\sim$$264$ hours of wall-clock time to generate a single trajectory over $10^9$ steps (on NVIDIA GeForce 
GTX 295 graphics card). The memory demand for Ran$2$ is quite high ($>$$250MB$ for $N$$=$$10^6$). In addition, 
implementing Ran$2$ on the GPU does not lead to a substantial speedup compared to the CPU-based implementation (Fig.~5). 
By contrast, the Hybrid Taus, and Lagged Fibonacci RNGs perform almost equally well in terms of the computational 
time and memory usage (Fig.~5). These generators require a small amount of memory ($<$$15$$-$$20MB$) even for a 
large system of $10^6$ particles (data not shown).


\section{Discussion and Conclusion}
\label{Discussion}

Increasing the computational speed of a single CPU core becomes more and more challenging for CPU manufacturers. 
With accelerated working frequency of modern CPUs, high power throughput results in CPU overheating, which prohibits 
unlimited growth in their computational power. In this regard, graphics processors are emerging as an alternative 
type of computing devices that evolve through increasing the number of computational cores rather than working 
frequency of a few cores. The highly parallel architecture of the GPU device provides an alternative computational 
platform that allows one to utilize multiple ALUs on a single processor. However, this comes at a price of 
having smaller cache memory and reduced flow control. Hence, to harvest raw computational power offered by the GPU 
in a particular application, one has to re-design computational algorithms that have been used on the CPU for several 
decades. The programmer has to be able to decompose each computational task into many independent threads of execution. 
In addition, care has to be taken to ensure coalescent memory access, and proper threads synchronization and communication.

Random number generators (RNGs) are needed for most of computer applications such as simulations of stochastic systems, 
probabilistic algorithms, and numerical analysis among others. We described the one-RNG-per-thread approach and the 
one-RNG-for-all-threads approach for random number generation on the GPU (Fig.~1), which we applied to the LCG, and 
to the Ran$2$, Hybrid Taus, and Lagged Fibonacci generators. We have tested these RNGs using Langevin 
simulations of $N$ independent Brownian particles, evolving on the harmonic potential. The LCG, Hybrid Taus, and 
Ran$2$ algorithms were realized on the GPU as independent RNGs producing many streams of random numbers 
at the same time (one-RNG-per-thread approach, Fig.~2). Additive Lagged Fibonacci algorithm was implemented using
many threads generating a single sequence of random numbers (one-RNG-for-all-threads approach, Fig.~3). The Hybrid Taus 
and Lagged Fibonacci algorithms of good statistical quality \cite{GPUGems3,Lecuyer07,Brent92} provide random numbers 
at a computational speed almost equal to that of the Quick and Dirty LCG (Fig.~4), and the associated memory demand 
is rather low (Figs.~5). Their long periods are sufficient to describe stochastic dynamics of a very large system 
($N$$>$$10^6$ particles) on a long timescale ($n$$>$$10^9$ simulation steps). This makes the Hybrid Taus and Lagged 
Fibonacci algorithms a very attractive option for molecular simulations of biomolecules on the GPU. Ran$2$ is a 
well tested generator of proven statistical quality (Fig.~4) \cite{Recipes}. It is probably the best RNG choice for 
molecular simulations on the CPU, but it works almost ten-fold slower on the GPU and requires large memory area (Fig.~5). 
Because using Ran$2$ in the molecular simulations of large systems can decrease significantly the computational 
speed of numerical modeling, Ran$2$ can be used in the simulations of small systems ($N$$\le$$10^3$ particles). 

Statistical characteristics of random numbers, generated by using the one-RNG-per-thread approach, do not improve 
with the increasing system size. In this setting, each RNG working in each thread uses its own state and, hence, 
increasing the number of threads (number of particles $N$) results in the increased number of generators, but it does 
not improve their statistical qualities. In the one-RNG-for-all-threads approach, streams of random numbers 
are produced in many threads running in parallel and sharing the same state variables. As a result, statistical 
properties of the random numbers improve with the increasing size of the RNG state. This is a general property 
of RNG implementation based on the one-RNG-for-all-threads approach \cite{Lecuyer07,Brent92}. For this reason, we 
recommend the Lagged Fibonacci RNG for compute-intensive LD simulations and MD simulations in implicit solvent of 
large biomolecules, and in parallel tempering algorithms including variants of the replica exchange method. Also, 
in the one-RNG-for-all-threads approach only one sequence on random numbers is generated, which makes is possible 
to compare directly the results of simulations on the CPU and on the GPU. This can be used in benchmark tests to 
estimate numerical errors due to single precision floating point arithmetic, rounding-off errors, or to identify 
bad memory reads on the GPU. 

Profiling the computational performance of the Hybrid Taus and Lagged Fibonacci generators have revealed that for
these RNGs the execution time scales sublinearly with $N$ (i.e. remains roughly constant) for $N$$<$$5$$\times$$10^3$ 
due to insufficient parallelization, but grows linearly with $N$ for larger sistems when all ALUs on the GPU become fully 
loaded (Fig.~6). Analysis of the execution time for Hybrid Taus and Lagged Fibonacci (RNG time) with the 
time of generation of deterministic dynamics, i.e. without the Gaussian random forces (dynamics without RNG time), 
shows that it takes slightly longer to generate random numbers than to propagate the dynamics to the next time step 
(Fig.~6). This is a pretty high performance level rendering the fact that the potential energy function used in 
our model simulations does not involve long-range interactions (Lennard-Jones type potential). Using the Hybrid 
Taus and Lagged Fibonacci RNG leads to a substantial $25$$-$$35$-fold speedup, as compared to the CPU-based implementation 
of these generators within the same LD algorithm. Given higher statistical quality of the Hybrid Taus and Lagged 
Fibonacci RNGs, these generators is a reasonable choice for the GPU-based implementations of molecular simulations 
(Fig.~5). Hybrid Taus allows one to obtain faster acceleration, compared to Lagged Fibonacci, but the latter has 
an important advantage over the former, namely, that it can be ported to new graphics cards that utilize Multiple 
Instruction Multiple Data (MIMD) architecture \cite{Fermi}. We also applied stringent statistical tests of randomness 
to access the statistical properties of random numbers produced by using our GPU-based implementation of the Lagged 
Fibonacci RNG. We found that even when a small short lag $sl$$=$$1252$ is used, this RNG does not fail a single 
tests in the DIEHARD test suite \cite{Marsaglia96}, and passes the BigCrush battery of tests in the TestUO1 package 
\cite{Lecuyer07}. 

In conclusion, the development of new Fermi architecture (NVIDIA) \cite{Fermi} and Larrabee architecture (Intel) 
\cite{Larrabee}, both equipped with $512$ ALUs, is an important next step for general purpose GPU computing. 
These next generation processors will utilize MIMD protocol, which will enable a developer to use many ALUs in 
independent computations so that different cores can perform concurrently different computational procedures 
on multiple data sets. Also, high speed interconnection network will provide a fast interface for threads 
communication. These advances in computer architecture will enable the programmer to distribute a computational 
workload among many cores on the GPU more efficiently, and to reach an even higher performance level. In a context 
of MD simulations in implicit solvent and in LD simulations, it will become possible to compute random forces using 
much needed threads synchronization over the entire processor. This makes the one-RNG-for-all-threads approach, 
where thread synchronization is utilized, all the more relevant as it will allow one to obtain additional acceleration 
on the GPU device gaining from high speed threads communication. Importantly, the GPU-based implementation of the 
Lagged Fibonacci RNG, developed here, could be ported to new graphics processors with a few minor modifications. 
In addition, the one-RNG-for-all-threads approach to random number generation can also be used to develop GPU-based 
implementations of the Mersenne Twister RNG, one of the most revered generators \cite{Matsumoto92,Matsumoto94,Matsumoto98}, 
and several other generators, including multiple recursive (MRG) and linear/generalized shift feedback register 
(LSFR/GSFR) generators, such as $4$-lag Lagged Fibonacci algorithm \cite{Lecuyer07,Marsaglia99}. Work in this 
direction is in progress.


\bigskip
\noindent {\bf Acknowledgements:} Acknowledgement is made to the donors of the American Chemical Society 
Petroleum Research Fund (grant PRF $\#$$47624$$-$$G6$) for partial support of this research (VB). This 
work was also supported in part by the grant ($\#$$09$$-$$07$$12132$) from the Russian Foundation for 
Basic Research (VB, YK and AZ).


\appendix
\label{Appendix}
\renewcommand{\thetable}{\Roman{table}}

\section{One-RNG-per-thread approach: Hybrid Taus and Ran$2$}
\label{Appendix.ManyRNGs}


\noindent
In the pseudocodes, that describe the GPU-based implementation of Hybrid Taus and Ran$2$ RNGs, superscript 
$h$ is used to denote the host (CPU) memory, whereas superscript $d$ indicates data stored in the 
device (GPU) global memory. Also, a section of the code executed on the GPU is in the same listing as the code 
for the CPU. In CUDA implementations, the corresponding code for the GPU device is in a separate 
kernel.

\noindent
{\bf{Algorithm 1:}} Hybrid Taus algorithm.
\algsetup{
	indent=2em,
	linenodelimiter=.
}
\begin{algorithmic}[1]
\REQUIRE $y_1^h[N]$, $y_2^h[N]$, $y_3^h[N]$ and $y_4^h[N]$ allocated in CPU memory
\REQUIRE $y_1^d[N]$, $y_2^d[N]$, $y_3^d[N]$ and $y_4^d[N]$ allocated in GPU global memory
\STATE $y_1^h[1\ldots N]$ to $y_4^h[1\ldots N] \leftarrow$ initial seeds 
\STATE $y_1^h[1\ldots N]$ to $y_4^h[1\ldots N] \rightarrow y_1^d[1\ldots N]$ to $y_4^d[1\ldots N]$ \COMMENT{copying initial seeds to GPU}
\GPU{}
\STATE $j_{th} \leftarrow$ thread index
\STATE $y_1$, $y_2$, $y_3$ and $y_4 \leftarrow y_1^d[j_{th}]$, $y_2^d[j_{th}]$, $y_3^d[j_{th}]$ and $y_4^d[j_{th}]$ \COMMENT{loading the state}
\FOR[generating four random numbers]{$i=1$ to $4$; $i++$} 
\STATE $b \leftarrow (((y_1 \ll c_{11}) \;\mathrm{XOR}\;y_1) \gg c_{21})$
\STATE $y_1 \leftarrow (((y_1 \;\mathrm{AND}\; c_1) \ll c_{31})\;\mathrm{XOR}\;b$
\STATE $b \leftarrow (((y_2 \ll c_{12}) \;\mathrm{XOR}\;y_2) \gg c_{22})$
\STATE $y_2 \leftarrow (((y_2 \;\mathrm{AND}\; c_2) \ll c_{32})\;\mathrm{XOR}\;b$
\STATE $b \leftarrow (((y_3 \ll c_{13}) \;\mathrm{XOR}\;y_3) \gg c_{23})$
\STATE $y_3 \leftarrow (((y_3 \;\mathrm{AND}\; c_3) \ll c_{33})\;\mathrm{XOR}\;b$
\STATE $y_4 \leftarrow ay_4+ c$
\STATE Output $mult\times(\;\mathrm{XOR}\;y_1\;\mathrm{XOR}\;y_2\;\mathrm{XOR}\;y_3\;\mathrm{XOR}\;y_4)$ 
\ENDFOR \COMMENT{generating next random number}
\STATE $y_1$, $y_2$, $y_3$ and $y_4 \rightarrow y_1^d[j_{th}]$, $y_2^d[j_{th}]$, $y_3^d[j_{th}]$ and $y_4^d[j_{th}]$ \COMMENT{saving the current state}
\ENDGPU{}
\end{algorithmic}
In this listing, $b$ is a temporary unsigned integer variable, $y_1$, $y_2$, $y_3$, and $y_4$ are unsigned 
integer random seeds for three Tausworthe generators (lines $6$$-$$11$) and one LCG (line $12$). $\mathrm{XOR}$ 
is a binary operation of exclusive disjunction and ``$\gg$'' and ``$\ll$'' denote binary shift to the right and 
to the left, respectively. In the pseudocode, $mult$$=$$2.3283064365387$$\times$$10^{-10}$ is a multiplier that 
converts a resulting integer into a floating point number, $c_{11}$$=$$13$, $c_{21}$$=$$19$, $c_{31}$$=$$12$, 
$c_{21}$$=$$2$, $c_{22}$$=$$25$, $c_{23}$$=$$4$, $c_{31}$$=$$3$, $c_{32}$$=$$11$, $c_{33}$$=$$17$, $c_1$$=$$4294967294$, 
$c_2=4294967288$, and $c_3$$=$$4294967280$ are constant parameters for three Tausworthe generators \cite{Lecuyer96}, 
and $a$$=$$1664525$ and $c$$=$$1013904223$ are constant parameters for the LCG \cite{Recipes}. 

\noindent
{\bf{Algorithm 2:}} Ran$2$ algorithm. 
\begin{algorithmic}[1]
\REQUIRE $idum^h[N]$, $idum2^h[N]$, $iy^h[N]$ and $iv^h[N*NTAB]$ allocated in CPU memory
\REQUIRE $idum^d[N]$, $idum2^d[N]$, $iy^d[N]$ and $iv^d[N*NTAB]$ allocated in GPU global memory
\STATE $idum^h[1\ldots N] \leftarrow$ initial seeds
\FOR[loading all $N$ generators]{$i=0$ to $N-1$; $i++$}
\STATE $idum2^h[i] \leftarrow idum^h[i]$
\FOR{$j=NTAB+7$ to $0$; $j--$}
\STATE $k \leftarrow idum^h[i]/IQ1$
\STATE $idum^h[i] \leftarrow IA*(idum^h[i] - k*IQ1) - k*IR1$
\IF{$idum^h[i] < 0$}
\STATE $idum^h[i]=idum^h[i]+IM1$
\ENDIF
\IF{$j<NTAB$}
\STATE $iv^h[i*NTAB+j] = idum^h[i]$
\ENDIF
\ENDFOR
\ENDFOR \COMMENT{all $N$ generators are intialized}
\STATE $idum^h \rightarrow idum^d$; $idum2^h \rightarrow idum2^d$; $iy^h \rightarrow iy^d$; $iv^h \rightarrow iv^d$ 
\COMMENT{copying to GPU}
\FOR[starting simulation for $S$ steps]{$t=0$ to $S$; $t++$}
\GPU{}
\STATE $j_{th} \leftarrow$ thread index
\STATE $idum \leftarrow idum^d[j_{th}]$; $idum2 \leftarrow idum2^d[j_{th}]$; $iy \leftarrow iy^d[j_{th}]$ \COMMENT{copying to GPU local memory}
\STATE $x[4]$ \COMMENT{output vector for four random numbers}
\FOR[generating four random numbers]{$i=0$ to $4$; $i++$}
\STATE $k\leftarrow idum/IQ1$; $idum \leftarrow IA1*(idum-k*IQ1)-k*IR1$
\IF{$idum < 0$}
\STATE $idum=idum+IM1$
\ENDIF
\STATE $k\leftarrow idum2/IQ2$; $idum2 \leftarrow IA2*(idum2-k*iQ2)-k*IR2$
\IF{$idum2 < 0$}
\STATE $idum2=idum2+IM2$
\ENDIF
\STATE $j \leftarrow iy/NDIV$
\STATE $iv \leftarrow iv^d[j_{th}*NTAB+j]$ \COMMENT{portion of the RNG state in GPU global memory}
\STATE $iy = iv-idum2$; $idum \rightarrow iv^d[j_{th}*NTAB+j]$
\IF{$iy < 1$}
\STATE $iy \leftarrow iy+IMM1$
\ENDIF
\STATE $tempran \leftarrow AM*iy$
\IF{$tempran > RNMX$}
\STATE $x[i] \leftarrow RNMX$
\ELSE
\STATE $x[i] \leftarrow tempran$
\ENDIF
\ENDFOR \COMMENT{generating next random number}
\STATE $idum \rightarrow idum^d[j_{th}]$; $idum2 \rightarrow idum2^d[j_{th}]$; $iy \rightarrow iy^d[j_{th}]$ 
\COMMENT{saving to GPU global memory}
\STATE Output: $x$
\ENDGPU{}
\ENDFOR \COMMENT{next simulation step}
\end{algorithmic}

Once $N$ RNGs are initialized on the CPU (lines $1$$-$$14$), initial seeds for all generators are copied to the 
GPU global memory (line $15$). The GPU-based computations start on line $16$. Each thread locates the 
values of the RNG state in the GPU global memory using thread index and copies the values of variables $idum$, 
$idum2$ and $iy$ to the GPU local memory (line $18$). Array $iv$ is accessed via GPU global memory calls (lines 
$30$ and $31$). Each thread generates four random numbers (cycle starting on line $20$) and saves them to array 
$x[4]$. Current RNG state variables are updated in the GPU global memory (line $42$).


\section{One-RNG-for-all-threads approach: Lagged Fibonacci}
\label{Appendix.OneRNG}

\noindent
{\bf{Algorithm 3:}} Additive Lagged Fibonacci algorithm.
\begin{algorithmic}[1]
\REQUIRE $x^d[N]$ allocated in GPU global memory
\STATE $x^d[1\ldots ll] \leftarrow$ initial seeds
\FOR[starting simulations]{$t=0$ to $S$}
\GPU{}
\STATE $j_{th} \leftarrow$ thread index
\STATE $shift_0 \leftarrow (j_{th}+N*t)*RNS$
\FOR{$shift=shift_0$ to $shift_0+RNS-1$}
\STATE $x_{ll} \leftarrow x^d[shift\;\mathrm{mod}\;ll]$
\STATE $x_{sl} \leftarrow x^d[(shift+sl-ll)\;\mathrm{mod}\;ll]$
\STATE $x \leftarrow (x_{ll}\; op\; x_{sl})\;\mathrm{mod}\;m$
\STATE output $x$
\STATE $x \rightarrow x^d[shift\;\mathrm{mod}\;ll]$
\ENDFOR
\ENDGPU{}
\ENDFOR
\end{algorithmic}
To initialize a RNG, a CPU fills in $ll$ integer random seeds into $x^d$ arrays and copies them to the GPU (line $1$), 
where each thread computes the location ($shift0$) of an integer that corresponds to the location of the first random 
number to be produced. This is done using the current simulation step ($t$), thread index ($j_th$), the total 
number of threads ($N$), and the amount of random numbers needed at each step ($RNS$). Lines $6$$-$$10$ are repeated 
until $RNS$ random numbers are generated (cycle starting on line $5$) using addition operator $op$ (line $8$). For every 
random number, two integers from the RNG state have to be gathered (lines $6$ and $7$). These integers correspond to the 
long lag $ll$ and the short lag $sl$. Locations in the array of integers are modulo $ll$, which represents ``cycling'' 
through the array of state integers starting from the beginning of the array (when its end is reached). The resulting 
integer $x$ (line $8$) is reported (line $9$) and saved for the next steps (line $10$). When $RNS$ random numbers are 
needed in each thread at each step of a simulation, $sl$$>$$RNS$$\times$$N$ and $ll-sl$$>$$RNS$$\times$$N$ (total number 
of integers updated at each step is $RNS$$\times$$N$).


\bibliographystyle{ieeetr}
\bibliography{books,papers}


\newpage

\section*{\bf FIGURE CAPTIONS}

\noindent
{\bf Fig.~1.}
Flowchart for generation of random numbers using the one-RNG-per-thread approach (panel 
$a$) and the one-RNG-for-all-threads approach (panel $b$). In the one-RNG-per-thread setting, 
$N$ independent RNGs (for $N$ particles) are running concurrently in $N$ computational threads on the 
GPU device generating random numbers from the same sequence, but starting from different sets of 
initial seeds. Within the one-RNG-for-all-threads approach, a single RNG is used by all
$N$ threads running in parallel on the GPU sharing one set of seeds and producing $N$ subsequences of
the same sequence of random numbers. The computational workflow is indicated by the arrows, and $n$, 
$n+1$, $\ldots$ are the simulation steps.

\bigskip

\noindent
{\bf Fig.~2.}
GPU-based realization of the one-RNG-per-thread approach. The arrows represent the direction of computational 
workflow and data transfer. To launch an RNG on the GPU, $N$ sets of initial random seeds, one 
set per thread of execution (per particle) generated on the CPU, are transferred to the GPU global memory. 
Each thread reads corresponding seeds from the GPU global memory, and generates random numbers for just 
one step of a simulation using a particular RNG algorithm. When all random numbers have been produced at
the $n$-th step, each thread saves its RNG state to the GPU global memory so that it could be used at the 
next step ($n+1$).

\bigskip

\noindent
{\bf Fig.~3.} 
GPU-based realization of the one-RNG-for-all-threads approach and parallel implementation of the Lagged Fibonacci 
algorithm using the cycle division paradigm. The state of the Lagged Fibonacci RNG is represented by the circle of 
$ll$ integers. Initial seeds are generated on the CPU and copied to the GPU global memory. Generation of $N$ 
random numbers is done simultaneously in $N$ threads using Eq.~(\ref{eq3}) (shown by arrows). The obtained 
random numbers are saved to update the RNG state for future use. A grid of computational threads is moving 
along the same sequence of random numbers, each time rewriting $N$ state variables that appear $ll$ positions 
earlier in the sequence. The dark grey squares represent the state variables, and the updated portion of the 
RNG state at a given step; the black ``zero line'', which denotes the position of the first thread, shifts 
forward by $N$ positions at every next step.

\bigskip 

\noindent
{\bf Fig.~4.}
Semilogarithmic plots of the average particle position $\langle X(t)\rangle$ (panels $a$ and $b$) 
and two-point correlation function $C(t)$$=$$\langle X(t)X(0)\rangle$ (panel $c$) for a system of $N$ 
harmonic oscillators in a stochastic thermostat. Theoretical curves of $\langle X(t)\rangle$ and $C(t)$ are 
compared with the results of Langevin simulations obtained using the LCG, Hybrid Taus, Ran$2$, 
and Lagged Fibonacci algorithms. Equilibrium fluctuations of $\langle X(t)\rangle$ on a longer timescale,
obtained using LCG, are magnified in panel $b$, where one can observe a repeating pattern due to the
inter-stream correlations among $N$ streams of random numbers.

\bigskip

\noindent
{\bf Fig.~5.}
Computational performance of the GPU-based implementations of the LCG, Ran$2$, Hybrid Taus, and Lagged 
Fibonacci algorithms in LD simulations of $N$ three-dimensional harmonic oscillators in a stochastic
thermostat (color code is explained in the graphs). Panel $a$: A logarithmic plot of the execution time (per 
$10^3$ steps) as a function of the system size $N$. Threads have been synchronized on the CPU at 
the end of each step to imitate the LD simulations of a biomolecule. As a reference, also shown is the simulation 
time with the Ran$2$ algorithm implemented on the CPU. Panel $b$: Memory demand, i.e. the amount of memory needed for 
an RNG to store its current state, as a function of $N$. A step-wise increase in the memory usage for Lagged 
Fibonacci at $N$$\approx$$0.6$$\times$$10^5$ is due to change in the values of constant parameters (Table~I).

\bigskip

\noindent
{\bf Fig.~6.}
Computational time (per $10^3$ steps) for an end to end application of LD simulations of $N$ three-dimensional 
harmonic oscillators in a stochastic thermostat using the Hybrid Taus (panel $a$) and Lagged Fibonacci algorithm 
(panel $b$), as a function of $N$ (color code is explained in the graphs). The simulation time for the full LD 
algorithm (Langevin Dynamics) is compared with the time for generating random numbers (RNG) and with the time 
required to obtain deterministic dynamics without random numbers (Dynamics w/o RNG). The computational speedup 
(CPU time/GPU time) is displayed in the insets.


\newpage

\section*{}

\begin{table}
\label{Table1}
\parbox[t]{6.5in}
{\caption{Constant parameters, i.e. the short lag $sl$ and the long lag $ll$, for the Lagged Fibonacci RNG 
for molecular simulations of a system of size $N$ (taken from Ref.~\cite{Brent92,Brent03}).}}

\vspace{.2in}
\small{
\begin{tabular}{| l | c | c | c | c | c | c | c | c | c | c | c | c |}\hline
$N$ & $<$$1252$  & $<$$3004$ & $<$$5502$ & $<$$10095$ & $<$$12470$ & $<$$23463$ & $<$$54454$ & $<$$279695$ & $<$$288477$ & $<$$1010202$  \\\hline
$sl$ & $1\,252$  & $3\,004$ & $5\,502$ & $10\,095$ & $12\,470$ & $23\,463$ & $54\,454$ & $279\,695$ & $288\,477$ & $1\,010\,202$ \\\hline
$ll$ & $2\,281$  & $4\,423$ & $9\,689$ & $19\,937$ & $23\,209$ & $44\,497$ & $132\,049$ & $756\,839$ & $859\,433$ & $3\,021\,377$ \\\hline
\end{tabular}
}

\end{table}


\begin{table}
\label{Table2}
\parbox[t]{6.5in}
{\caption{Memory usage (in bytes/thread), and the number of GPU global memory calls, i.e. the numbers of 
read/write operations per one random number ($M_1$) and for four random numbers ($M_2$), for generation of random 
numbers on the GPU at each step using the LCG, and the Hybrid Taus, Ran$2$, and Lagged Fibonacci RNGs. In molecular 
simulations, four random numbers are needed at each step to generate three ($x$, $y$, and $z$) components 
of the Gaussian random force per particle.}}

\vspace{.2in}
\begin{tabular}{| c | c | c | c | c |}\hline
Parameter & \quad LCG \qquad & \quad Hybrid Taus \qquad & \quad Ran$2$ \qquad & Lagged Fibonacci \quad \\\hline
\quad bytes/thread \qquad & $4$ & $16$ & $280$ & $12$ \\
$M_1$ & $1/1$ & $4/4$ & $4/4$ & $3/1$ \\
$M_2$ & $1/1$ & $4/4$ & $7/7$ & $12/4$ \\\hline
\end{tabular}

\end{table}


\newpage

\begin{figure}
\label{Fig1}
\centering
\includegraphics[width=7.8in,angle=90]{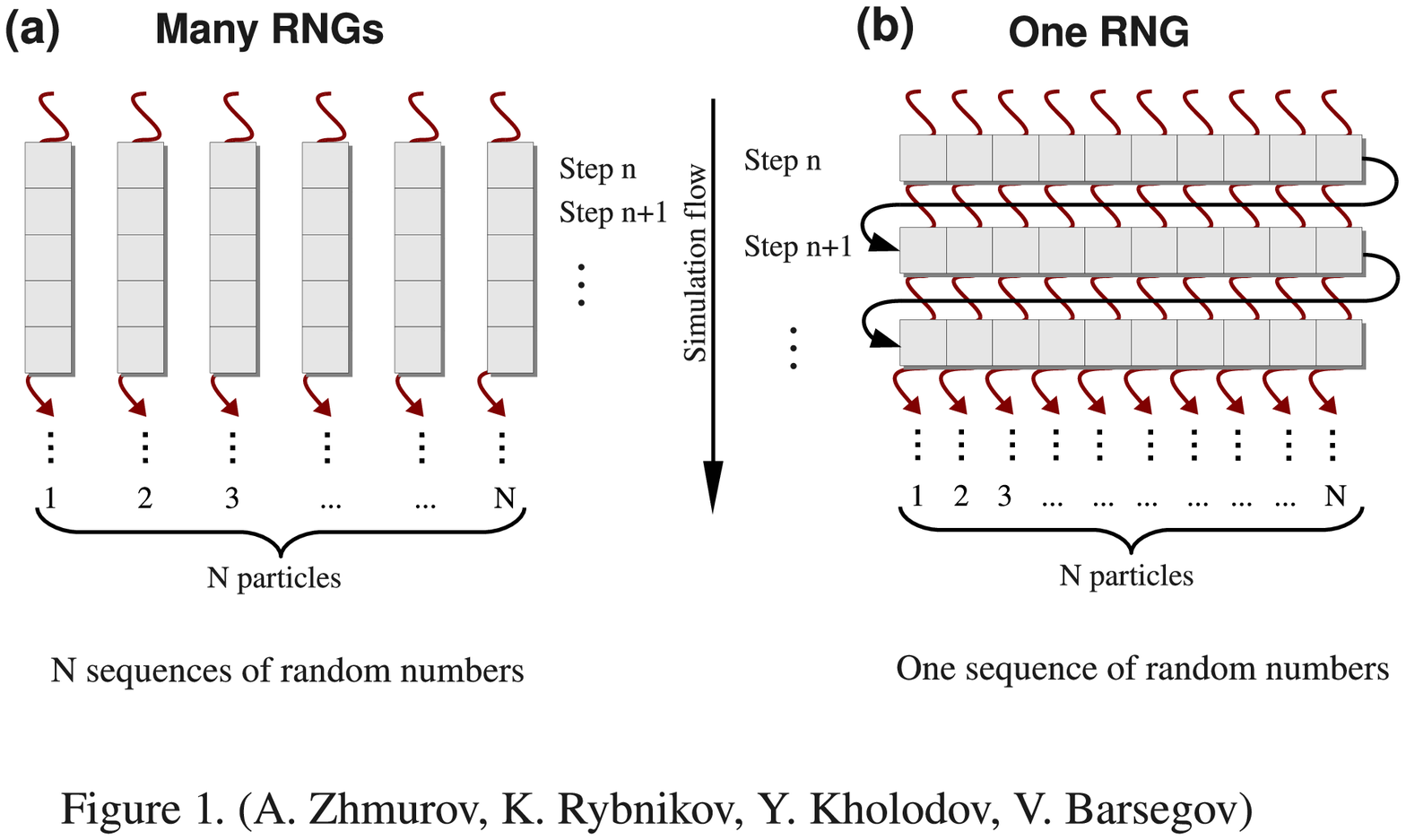}
\end{figure}

\newpage

\begin{figure}
\label{Fig2}
\includegraphics[width=7.8in,angle=90]{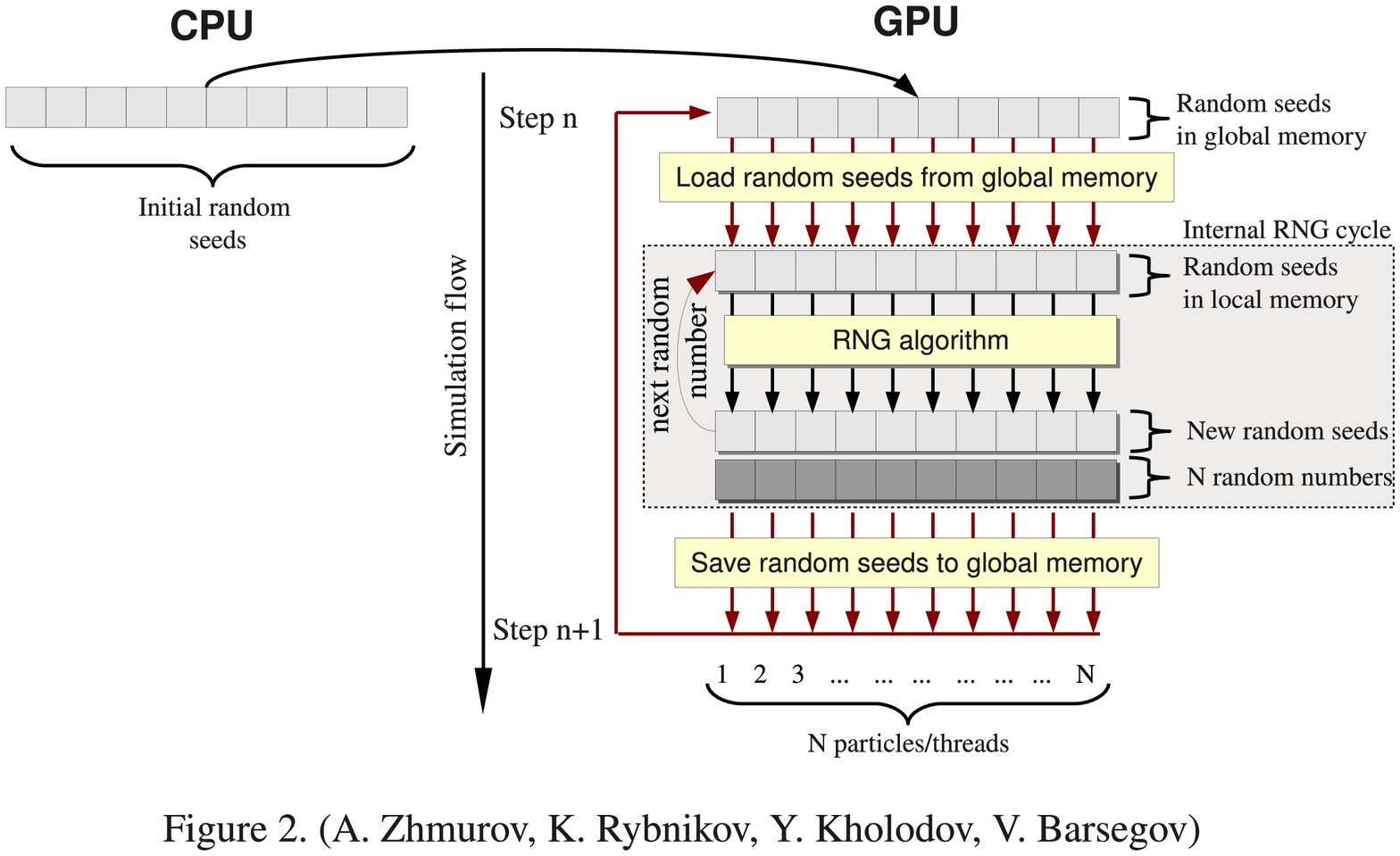}
\end{figure}

\newpage

\begin{figure}
\label{Fig3}
\includegraphics[width=6.0in]{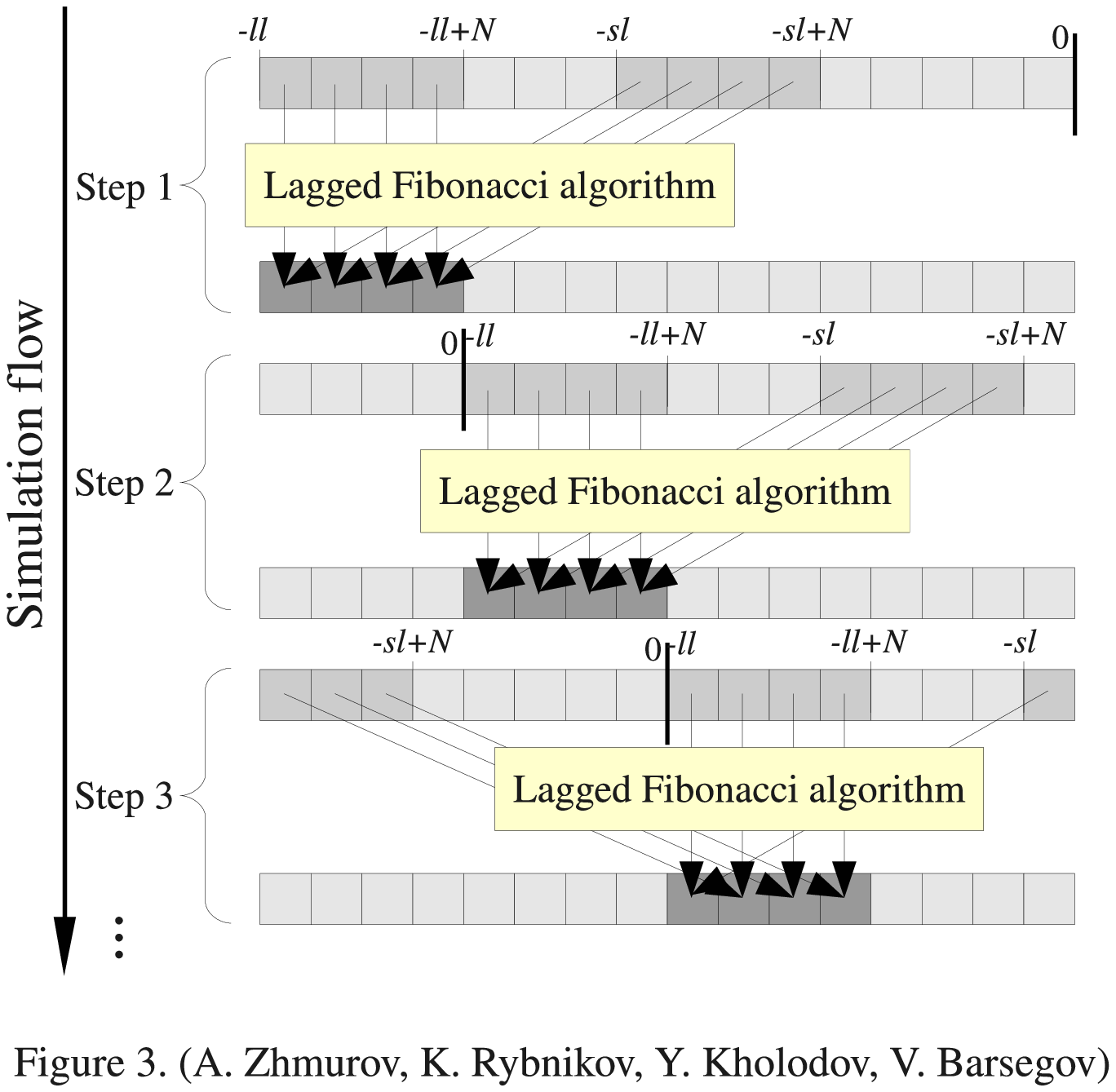}
\end{figure}

\newpage

\begin{figure}
\label{Fig4}
\includegraphics[width=3.5in]{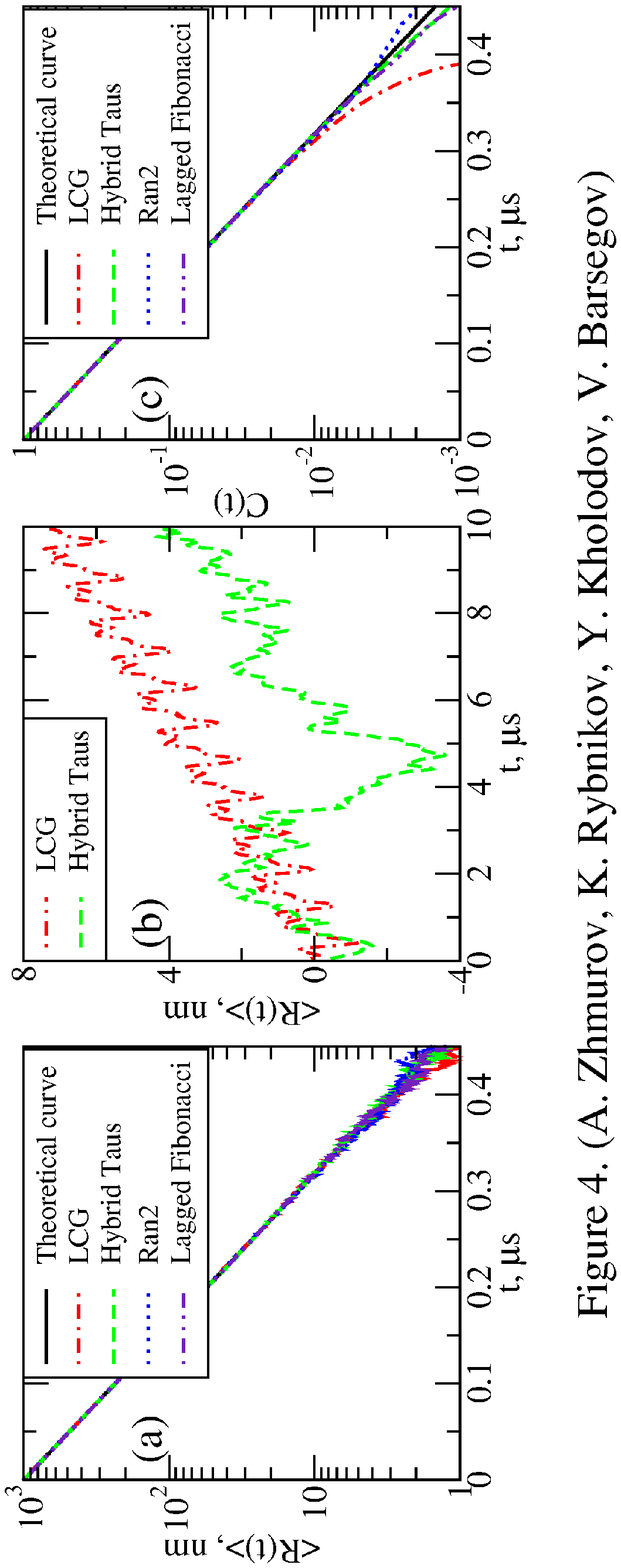}
\end{figure}

\newpage

\begin{figure}
\label{Fig5}
\includegraphics[width=4.0in]{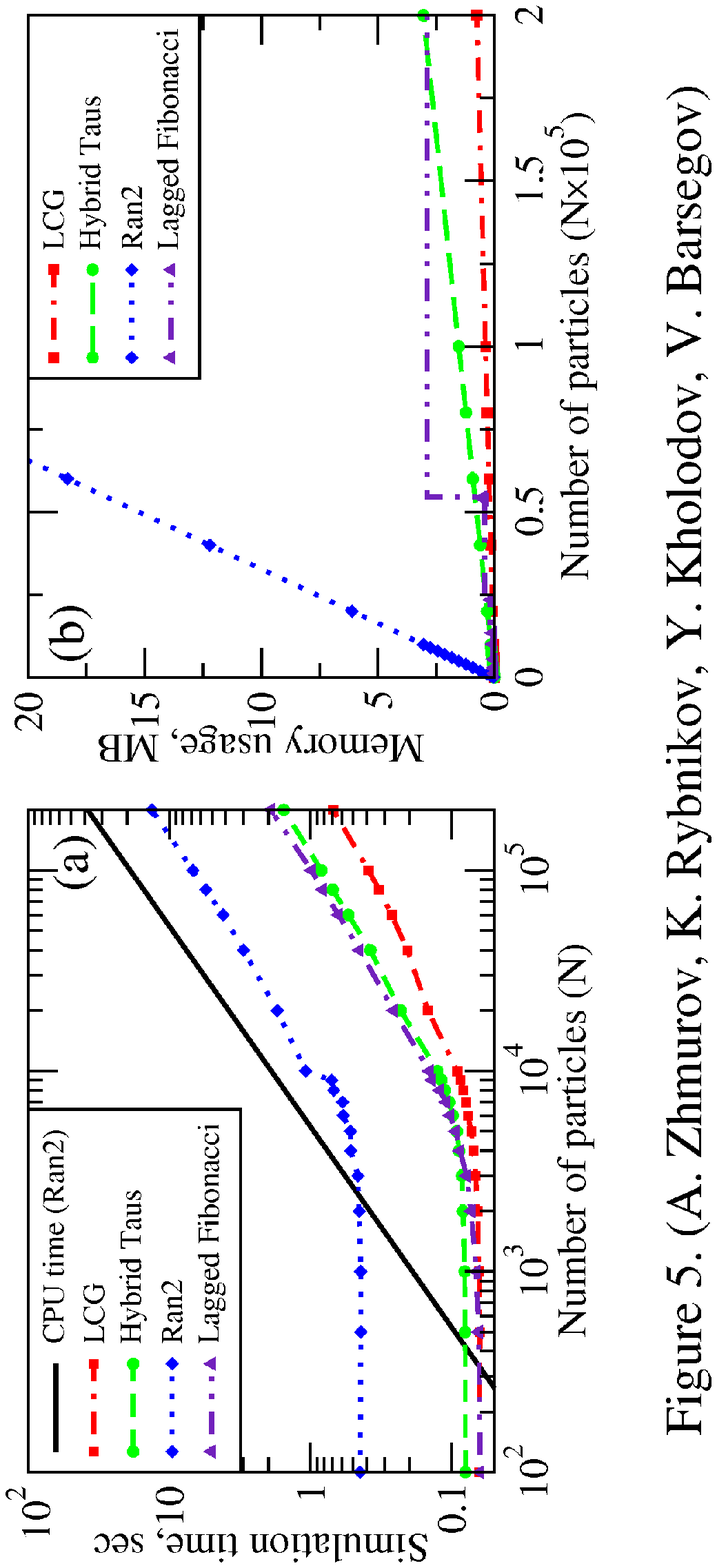}
\end{figure}

\newpage

\begin{figure}
\label{Fig6}
\includegraphics[width=4.0in]{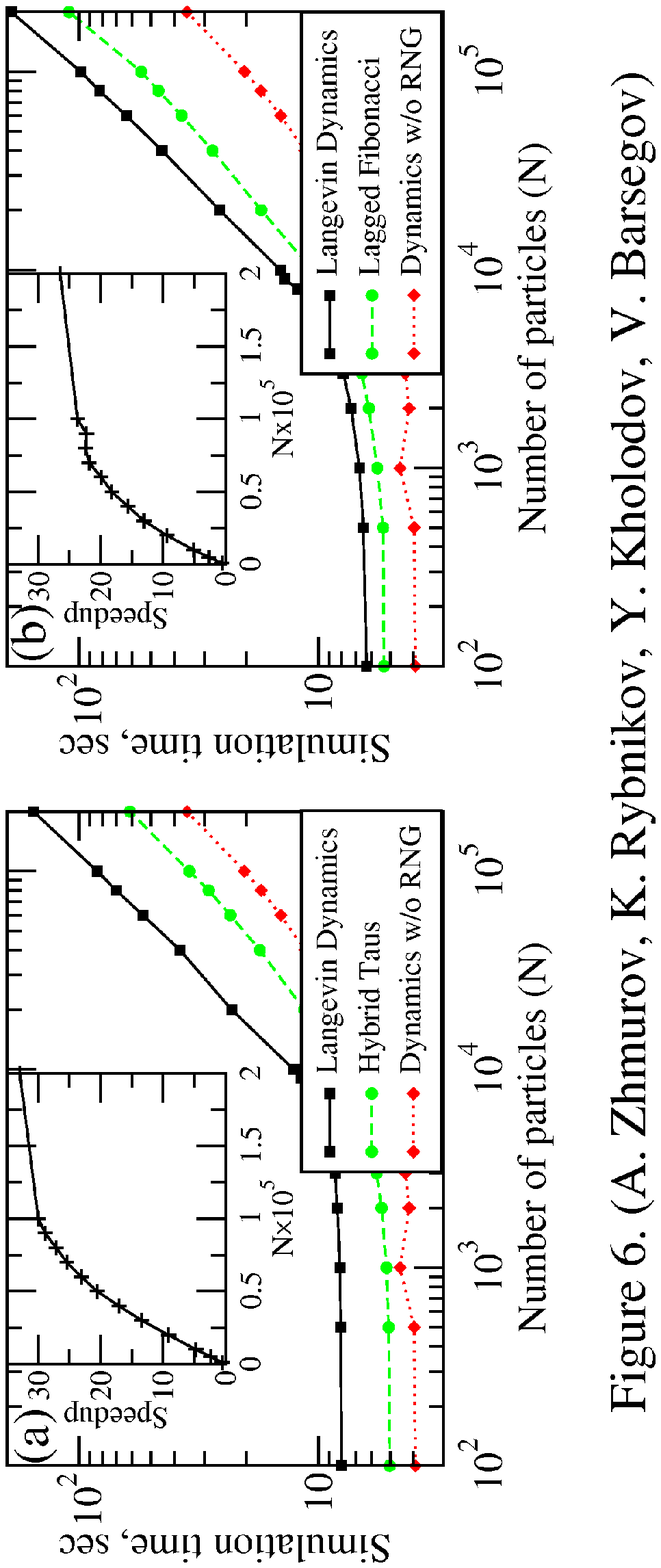}
\end{figure}


\end{document}